\documentstyle[12pt,fleqn]{article}
\catcode`\@=11
\begin{document} \openup6pt
\small
\title{\Large{Naked Singularities in Higher Dimensional Inhomogeneous
Dust Collapse}}

\author{S.G.~Ghosh\thanks{E-mail:
sgghosh@hotmail.com}   \\
 Department of Mathematics, Science College, Congress Nagar, \\
Nagpur - 440 012, India
\and and \\
A.~Beesham\thanks{Author to whom all correspondence should be
directed; E-mail: abeesham@pan.uzulu.ac.za} \\
Department of  Mathematical Sciences, University of Zululand, \\
Private Bag X1001, Kwa-Dlangezwa 3886, South Africa}

\date{}

\maketitle



\begin{abstract}
We investigate the occurrence and nature of a naked singularity in
the gravitational collapse of an inhomogeneous dust cloud described by a
non self-similar higher dimensional Tolman spacetime.  The necessary
condition for the formation of a naked singularity or a black hole is obtained.
  The naked singularities are found to be gravitationally strong
in the sense of Tipler and provide another example that
 violates the cosmic censorship conjecture.
\end{abstract}
{\bf PACS number(s):} 04.20.Dw
\bigskip
\section{Introduction}
The cosmic censorship hypothesis (CCH) \cite{rp} is an important
source of inspiration for research in general relativity.  It
states that the singularities produced by gravitational collapse
must be hidden behind an event horizon.  Moreover, according to
the strong version of the CCH, such singularities are not even locally
naked, i.e., no non-spacelike curve can emerge from such
singularities.   Since the conjecture has as  yet no precise
 mathematical formulation, a significant
amount of attention has been given to studying examples of gravitational
collapse which lead to naked singularities with matter content that
satisfies the energy conditions (see \cite{r1} for recent
reviews).  In particular, the collapse of spherical
matter in the form of dust forms naked shell focusing
singularities violating the CCH \cite{es}-\cite{djd}.

Recently, significant efforts have been expended to study gravitational
collapse models in higher dimensional spacetime.  llha {\it et al}
\cite{hd} have
constructed Oppenheimer-Snyder models in higher dimensional spacetime.
The self-similar  solution of
 spherically symmetric gravitational collapse of a scalar field in higher
dimensions is obtained in \cite{af}.  Gravitational collapse of a perfect fluid
in higher dimensional spacetime is studied by Rocha and Wang \cite{rw}.
In particular, one would like to
 understand the role played by extra dimensions in the formation and the
nature of singularities.
The solution for the higher dimensional spherically symmetric
dust collapse is obtained in \cite{bsc}, which  reduces to the
well known Tolman-Bondi solution when the dimension of the spacetime becomes four.
We shall call it the higher dimensional Tolman spacetime.
They have shown the occurrence of naked singularities for the self-similar
case.
However, self similarity is a strong geometric condition on the spacetime
and thus gives rise to a possibility that the naked singularity could be
the artifact of a geometric condition rather than the gravitational
 dynamics of matter therein.

The objective of this paper is to analyze in some detail the collapse of
an inhomogeneous dust cloud in higher dimensions to cover both non self-similar
and self-similar models.  We also access the curvature strength of central
shell focusing singularities.
We find that gravitational collapse of a  non self-similar higher dimensional
spacetime gives rise to a naked strong-curvature shell-focusing
singularity, providing an explicit counter-example to the CCH.

\section{Higher Dimensional Tolman Solution}
The idea that spacetime should be extended from four to
higher dimensions was introduced by Kaluza and Klein \cite{kk} to
unify gravity and electromagnetism.
Five-dimensional ($5D$) spacetime is  particularly more relevant because
both $10D$ and $11D$ supergravity theories yield  solutions where a  $5D$
spacetime results after dimensional reduction \cite{js}.
The metric for the $5D$ case, in
comoving coordinates coordinates, assumes the form:
\begin{equation}
ds^2 = -dt^2 + \frac{R'^2}{1+f(r)} dr^2 +
R^2 d \Omega^2 \label{eq:me}
\end{equation}
where $d\Omega^2 = d \theta_1^2+
sin^2 \theta_1 (d \theta_2^2+sin^2 \theta_2 d\theta_3^2)$
is the metric of a $3$ sphere,
 $r$ is the comoving radial coordinate, $t$ is the proper time of freely
falling shells, $R$ is a function of $t$ and $r$ with $R>0$
 and a prime denotes a partial derivative with respect to r.
The energy momentum tensor is of the form:
\begin{equation}
 T_{ab} =  \epsilon(t,r) u_{a}u_{b} \label{eq:emt}
\end{equation}
where $u_a$ is the five velocity.  The function $R(t,r)$ is the solution of
\begin{equation}
\dot{R}^2 = \frac{F(r)}{R^2} + f(r) \label{eq:fe}
\end{equation}
where an overdot denotes the partial derivative with respect to
t. The functions  $F(r)$ and  $f(r)$ are arbitrary, and  result
from the integration of the field equations. They are referred
to as  the mass and energy functions, respectively.  Since in
the present discussion we are concerned with gravitational
collapse, we require that $\dot{R}(t,r) < 0$.

The energy density $\epsilon(t,r)$ is given by
\begin{equation}
 \epsilon(t,r) = \frac{3 F'}{2 R^3 R'}  \label{eq:edt}
\end{equation}
We have used units which fix the speed of light and the gravitational constant
via $8\pi G = c^4 = 1$.  For physical reasons, one assumes that
the energy density $\epsilon$ is everywhere nonnegative $(\geq 0)$.
Eq. (\ref{eq:fe}) can  easily be
integrated to
\begin{equation}
t - t_{c}(r) = - \frac{R^2}{\sqrt{F}}G(fR^2/F)  \label{eq:ct}
\end{equation}
where $G(x)$ is the function given by
\begin{equation}
G(x) = \left\{ \begin{array}{ll}
        \frac{\sqrt{1+x}}{x},           &       \mbox{$ x \neq 0$}, \\
        \frac{1}{2}                     &       \mbox{$x = 0 $}.   \\
                \end{array}
        \right.                         \label{eq:fg}
\end{equation}
and where $t_{c}(r)$ is a function of integration which represents the time
taken by the shell with coordinate $r$  to collapse to the centre. This is
unlike the $4D$
case, where the functional form of $G$ is rather complicated \cite{rn,dj}.
As it is possible
to make an arbitrary relabeling of spherical dust shells by
$r \rightarrow g(r)$, without loss of generality, we fix the labeling by
requiring that, on the hypersurface $t = 0$, $r$ coincide with the
radius
\begin{equation}
R(0,r) = r              \label{eq:ic}
\end{equation}
This corresponds to the following choice of $t_{c}(r)$
\begin{equation}
t_{c}(r) =  \frac{r^2}{\sqrt{F}} G({fr^2}/{F})  \label{eq:ct1}
\end{equation}
We denote by $\rho(r)$ the initial density:
\begin{equation}
\rho(r) \equiv \epsilon(0,r) = \frac{3 F'}{2 r^3}
\Rightarrow F(r) = \frac{2}{3} \int \rho(r) r^3 dr  \label{eq:Fr}
\end{equation}
Given a regular initial surface, the time for the occurrence of the central shell-focussing singularity for the collapse developing from that surface is reduced as compared to the $4D$ case for the marginally bound collapse. The reason for this stems from the form of the mass function in Eq. (\ref{eq:Fr}). In a ball of radius $0$ to $r$, for any given initial density profile $\rho(r)$, the total mass contained in the ball is greater than in the corresponding $4D$ case. In the $4D$ case, the mass function $F(r)$ involves the integral $\int \rho(r) r^2 dr$ \cite{r1}, as compared to the factor $r^3$ in the $5D$ case. Hence, there is relatively more mass-energy collapsing in the spacetime as compared to the $4D$ case, because of the assumed overall positivity of mass-energy (energy condition). This explains why the collapse is faster in the $5D$ case.

The easiest way to detect a singularity in a spacetime is to
observe the divergence of some invariant of the Riemann tensor.
Next we calculate one  such quantity, the Kretschmann scalar ($
K = R_{abcd} R^{abcd}$, $R_{abcd}$ the Riemann tensor).  For the
metric (\ref{eq:me}), it reduces to
\begin{equation}
K = 7 \frac{F'^2}{R^6 R'^2} - 36 \frac{F F'}{R^7 R'} + 78 \frac{F^2}{R^8}
      \label{eq:ks}
\end{equation}
The Kretschmann scalar and energy density both diverge at $t=t_{c}(r)$
confirming the presence of a scalar polynomial curvature singularity \cite{he}.
Thus the time coordinate and radial coordinate are respectively
in the ranges $ - \infty < t < t_{c}(r)$ and $0 \leq r < \infty$.
It has been shown \cite{rn} that
shell crossing singularities (characterized by $R'=0$ and $R>0$) are
gravitationally weak and hence such singularities cannot be considered
 seriously.  Christodoulou \cite{dc} pointed out in the $4D$ case
that the non-central singularities are not naked.
Hence, we shall confine our discussion to the central shell focusing
singularity.

\section{Existence and Nature of Naked Singularity}
It is known that, depending upon the inhomogeneity factor, the $4D$ Tolman-Bondi
metric admits a central shell focusing naked singularity in the sense that
outgoing geodesics emanate from the singularity.  Here we wish to
investigate the similar situation in our higher dimensional spacetime.
In what follows, we shall confine ourselves to the marginally bound case $(f=0)$.
  Eq. (\ref{eq:ct}), by virtue of eq.
(\ref{eq:ic}), leads to
\begin{equation}
R^2 = r^2 - 2 \sqrt{F} t        \label{eq:sf}
\end{equation}
and the energy density becomes
\begin{equation}
\epsilon(t,r) = \frac{3/2}{ \left[t -  \frac{2 r \sqrt{F}}{F'} \right]
\left[t - \frac{r^2}{2 \sqrt{F}} \right]}
\end{equation}
We are free to specify $F(r)$ and we consider a
class of models which are non self-similar in general, and as a
special case, the  self-similar models can be constructed from them.
In particular, we suppose that $F(r) = r^2 \lambda(r)$ and
$\lambda(0) = \lambda_{0} > 0 $ (finite). With this choice of $F(r)$,
the density behaves as inversely proportional to the square of time at the
centre, and $F(r) \propto r^2$ in the neighborhood of $r=0$.
For spacetime to be self-similar, we require that $\lambda(r) = const.$
This class of models for $4D$ spacetime is discussed in refs. \cite{dj}.
From eq. (\ref{eq:edt}) it is seen that the density at the centre (r=0) behaves
with time as $\epsilon = 3/2t^2$.  This means that the density becomes singular
at $t=0$ and finite at any time $t = t_0 < 0$.
Thus the singularity arises from dust collapse which had
a finite density distribution in the past on an initial epoch. At this initial nonsingular epoch, all the physical parameters, including the density, are finite and well-behaved. Thus our collapse starts from regular initial data.

We wish to investigate if the singularity, when the central
shell with  comoving coordinate $(r=0)$ collapses to the centre
at time $t=0$, is naked.  The singularity is naked iff
there exists a null geodesic which emanates from the
singularity.  Let $K^a=dx^a/dk$ be the tangent vector to the radial null geodesic,
where $k$ is an affine parameter.  Then we derive the following
equations
\begin{equation}
\frac{dK^t}{dk} + \dot{R}' K^r K^t = 0 \label{eq:ngf}
\end{equation}
 \begin{equation}
\frac{dt}{dr} =  \frac{K^t}{K^r} = R' \label{eq:ng}
\end{equation}
The last equation, upon using eq. (\ref{eq:fe}), turns out to be
\begin{equation}
\frac{dt}{dr} = \frac{2 r - \frac{t F'}{\sqrt{F}}}{2 \sqrt{r^2 - 2t \sqrt{F}}}
 \label{eq:ng1}
\end{equation}
Clearly this differential equation becomes singular at $(t,r)=(0,0)$.
We now wish to put eq. (\ref{eq:ng1}) in a form that will be more useful for
subsequent
calculations. To this end, we define two new functions $\eta = rF'/F$ and
$P = R/r$.
From eq. (\ref{eq:sf}), for $f=0$, we have $\dot{R}=- \sqrt{F}/R$, and we can
express $F$ in terms of $r$ by $F(r)=r^{2} \lambda(r)$. Eq. (15) can thus be
re-written as
\begin{equation}
\frac{dt}{dr} = \left[ \frac{t}{r} \eta - 2 \sqrt{\lambda} \right] \dot{R}
= - \left[ \frac{t}{r} \eta - 2 \sqrt{\lambda} \right] \frac{\sqrt{\lambda}}
{P}  \label{eq:ng2}
\end{equation}
It can be seen that the functions $\eta(r)$
and $P(r,t)$ are well defined when the singularity is approached.

The nature (a naked singularity or a black hole) of the
singularity can be characterised by the existence of radial null
geodesics emerging from the singularity.  The singularity is
at least locally naked if there exist such geodesics, and
if no such geodesics exist, it is a black hole.
Let us  define $X = t/r$.  If the singularity is naked, then there
exists a real and positive value of $X_{0}$ as a solution to the
algebraic equation \cite{r1}
\begin{equation}
X_{0} = \lim_{t\rightarrow 0 \; r\rightarrow 0} X =
\lim_{t\rightarrow 0 \; r\rightarrow 0} \frac{t}{r}=
\lim_{t\rightarrow 0 \; r\rightarrow 0} \frac{dt}{dr} = R'      \label{eq:lm1}
\end{equation}
We insert eq. (\ref{eq:ng2}) into (\ref{eq:lm1}) and use the result
$\lim_{r\rightarrow 0} \eta = 2$ to get
\begin{equation}
X_{0} = \frac{2}{Q_0} \left[\lambda_{0} - X_{0} \sqrt{\lambda_{0}} \right]
\label{eq:pe1}
\end{equation}
where $Q(X) = P(X,0)$. From the definitions of $P$ and $X$, and
eq. (\ref{eq:sf}), we can derive the following equation
\begin{equation}
X - \frac{1}{2 \sqrt{\lambda}} = \frac{-P^2}{2 \sqrt{\lambda}} \label{eq:pe2}
\end{equation}
from which it is clear that $X \sqrt{\lambda}< 1/2$, as $P$ is a positive
function.  Since $Q(X)=P(X,0)$, from eq. (\ref{eq:pe2}) we get
$Q_{0} = \sqrt{1 - 2 X_{0} \sqrt{\lambda_{0}}}$.  Substituting this into
eq. (\ref{eq:pe1}) leads to the cubic equation
\begin{equation}
2z^3 + (4 \lambda_{0} - 1) z^2 - 8 \lambda_{0}^2 z + 4 \lambda_{0}^3 = 0
\label{eq:pe3}
\end{equation}
where $z = X_{0} \sqrt{\lambda_{0}}$.

We are interested in  positive
roots of  eq. (\ref{eq:pe3}) subject to the constraint that $z < 1/2$, in
which case outgoing null geodesics terminate at the singularity in the past.
  It is verified
numerically that that for $\lambda_{0} \leq 0.5480$ (correct to
four decimal places) eq. (\ref{eq:pe3}) has two positive real
roots which satisfy the constraint that $z = X_{0}
\sqrt{\lambda_{0}} < 1/2$.  For example, if $\lambda_{0}=0.25$, then eq.
(\ref{eq:pe3}) has the two roots $z = 0.1348$ and $0.4188$
(which correspond to two values $X_{0} = 0.2696$ and $0.8376$).
Thus it follows that the singularity will be at least locally naked when
$\lambda_{0} \leq 0.5480$.  On the other hand, if the inequality is
reversed, i.e., $\lambda_{0} > 0.5480$, no naked singularity occurs and
gravitational collapse of the dust cloud must result in  a
black hole.  In the analogous $4D$ case, one gets a quartic equation
and the shell focusing singularity is naked iff $\lambda_{0} <
0.1809$ \cite{dj}.  The global nakedness of the singularity can then be
seen by making a junction onto the higher dimensional Schwarzschild  spacetime, analogously to the $4D$ case (see \cite{bsc}). Jhingan and Magli \cite{r1} have pointed out that if locally naked singularities occur in dust spacetimes, then these spacetimes can be matched to spacetimes containing globally visible spacetimes.

\subsection{Self-Similar Case}
To support our analysis, we now specialise to the case of
self-similar spacetime, which has been analyzed earlier \cite{bsc} by a
different approach.
As already mentioned, for the spacetime to be self-similar, we
require that $F(r) = \lambda r^2$ ($\lambda = const.$) so that
\begin{equation}
R = r \sqrt{1 - 2 \sqrt{\lambda} \frac{t}{r}} \nonumber
\end{equation}
and $R'$ can be expressed in terms of the quantity $X = t/r$ as
\begin{equation}
R' = \frac{1 - \sqrt{\lambda}X}{\sqrt{1-2 \sqrt{\lambda}X}} \nonumber
\end{equation}
Eq. (\ref{eq:lm1}) with this results in:
\begin{equation}
2y^3 + (\lambda - 1) y^2 - 2 \lambda y + \lambda = 0 \label{eq:pe4}
\end{equation}
where $y = X \sqrt{\lambda}$.  Eq. (\ref{eq:pe4}) has positive roots,
subject to constraint that $y<1/2$, if $\lambda \leq 0.0901$.
(The two roots $y =0.2349$ and $0.4679$ of Eq. (\ref{eq:pe4}) correspond to
$\lambda = 0.05$). Thus referring to
our above discussion, self-similar collapse lead to a naked singularity
for  $\lambda \leq 0.0901$ and to formation of a black hole otherwise.
It is well known that the formation of a naked singularity can be understand
in terms of the inhomogeneity of the collapse.  If collapse is homogeneous
no naked singularity occurs.  On the other hand a naked singularity occurs
if the collapse is sufficiently inhomogeneous, i.e., if the outer shells collapse
much later than the central shells \cite{de}.  The parameter $B$ which gives
a measure of the inhomogeneity of the collapse is usually defined as $t_{c}(r) = Br$.
In our case, on comparing this with eq. (\ref{eq:ct1}) (with $f=0$),
$B = 1/2 \sqrt{\lambda}$.  So for the singularity to be naked we must have
$B > 1.6657$.  This is in  agreement with  earlier work \cite{bsc}.
\subsection{Strength of Naked Singularity}
An important aspect of a singularity is its gravitational strength
\cite{ft1}.
A singularity is gravitationally strong or simply strong if it destroys
by crushing or stretching any object which fall into it.  It is widely believed
 that a spacetime does not admit an extension through a singularity if it is
a strong curvature singularity in the sense of Tipler \cite{ft}.  Clarke and
Kr\'{o}lak \cite{ck} have shown that a sufficient condition for a strong
curvature singularity as defined by
Tipler \cite{ft} is that for at least one non-space like geodesic with affine
parameter $k$, in the limiting approach to the singularity, we must have
\begin{equation}
\lim_{k\rightarrow 0}k^2 \psi =
\lim_{k\rightarrow 0}k^2 R_{ab} K^{a}K^{b} > 0 \label{eq:sc}
\end{equation}
where $R_{ab}$ is the Ricci tensor.
Our purpose here is to investigate the above condition along future directed
radial null geodesics which emanate from the naked singularity. Now
$k^2 \psi$, with the help of eqs. (\ref{eq:emt})
can be expressed as
\begin{equation}
k^2 \psi = k^2 \frac{3F'(K^t)^2}{2R^3 R'}
= \frac{3F'}{2rP R'} \left[ \frac{k K^t}{R} \right]^2
\label{eq:sc1}
\end{equation}
Using our previous results, we find that
\begin{equation}
\lim_{k\rightarrow 0}k^2 \psi = \frac{3 \lambda_0 X_0}
{Q_0(X_0 - \sqrt{\lambda_0})^2} > 0
\end{equation}
as the singularity is naked for $X_{0} \sqrt{\lambda_{0}} < 1/2$.  Thus along
 radial null geodesics coming out from singularity, the strong curvature condition
is satisfied.

\section{Conclusion}
The occurrence and curvature strength of a shell focusing naked singularity
in a non self-similar higher dimensional spherically symmetric collapse
of a dust cloud has been investigated.  We found that naked singularities in
our case occur for a slightly higher value of the inhomogeneity parameter in
comparison to the analogous situation in the $4D$ case.  Along the null ray emanating
from the naked singularity, the strong curvature condition (\ref{eq:sc}) is
satisfied. The models constructed here are non self-similar
in general and for the special case $\lambda = const.$, reduce to the self-similar
case.  Whereas we have
applied this formalism to the $5D$ case, there is no reason to believe that
it cannot be extended to a spacetime of any dimension $(n \geq 4)$.
 The formation of
these naked singularities violates the strong CCH.  We do not
claim any particular physical significance to the $5D$ metric
considered.  Nevertheless we think that the results obtained here
have some interest in the sense that they do offer the
opportunity to explore  properties associated with naked
singularities.

\bigskip
\noindent
{\bf Acknowledgment:} SGG  would like to thank the University
of Zululand for hospitality, the NRF (South Africa) for
financial support, Science College, Congress Nagar, Nagpur
(India) for granting leave and  IUCAA, Pune for a visit where part of this
work was done.  AB thanks the Mehta Research
Institute, Allahabad, India for kind hospitality. The authors are grateful to an anonymous referee for constructive criticism.

\noindent

\small

\begin{thebibliography}{99}
\bibitem{rp} Penrose R 1969 {\it Riv del Nuovo Cimento} {\bf 1} 252
\bibitem{r1} Joshi P S 1993  {\it Global Aspects in Gravitation and
Cosmology} (Clendron Press, Oxford) \\
 Clarke C J S 1993
{\it Class. Quantum Grav.} {\bf 10}, 1375 \\ Wald R M 1997 {\it Preprint} gr-qc/9710068 \\ Jhingan S and Magli G 1999 {\it Preprint} gr-qc/9903103 \\ Joshi P S 2000 {\it Preprint} gr-qc/0006101
\bibitem{es} Eardley D M  and Smarr L 1979 {\it Phys. Rev.}
{\bf D 43} 2239
\bibitem{dc} Christodoulou D 1984 {\it Commun. Math. Phys.} {\bf 93} 171
\bibitem{de} Eardley D M 1986 {\it Gravitation in Astrophysics, NATO
 ASI Series}, eds B Carter and J B Hartle (Plenum Press, NY) pp 229-235
\bibitem{rn} Newman  R P A C 1986 {\it Class. Quantum Grav.} {\bf 3} 527
\bibitem{lz} Waugh B and Lake K 1988 {\it Phys. Rev.} {\bf D 38} 1315
\bibitem{jl} Lemos J P S 1991 {\it Phys. Lett.}
{\bf A 158} 271 \\ Lemos J P S 1992 {\it Phys. Rev. Lett.}  {\bf 68} 1447
\bibitem{dj} Dwivedi I H and Joshi P S 1992 {\it Class. Quantum Grav.}
 {\bf 9} L69
\\ Joshi P S and Singh T P 1995 {\it Gen. Rel. Grav.}
 {\bf 27} 921 \\  Joshi P S and Singh T P 1995 {\it Phys. Rev.}
 {\bf D 51} 6778
\bibitem{jd1} Joshi P S and Dwivedi I H 1993 {\it Phys. Rev.}
 {\bf D 47} 5357
\bibitem{djd} Deshingkar S S, Joshi P S and Dwivedi I H 1999
{\it Phys. Rev.} {\bf D 59} 044018
\bibitem{hd} llha A and  Lemos J P S 1997 {\it Phys. Rev. D} {\bf 55} 1788
\\ llha A,  Kleber A and Lemos J P S 1999 {\it J. Math. Phys.} {\bf 40}
3509
\bibitem{af} Frolov A V 1999 {\it Class. Quantum Grav.} {\bf 16} 407
\bibitem{rw} Rocha J F V and  Wang A  1999 {\it Preprint} gr-qc/9910109
 \bibitem{bsc} Banerjee A, Sil A and Chatterjee S 1994 {\it Astrophys. J.}
 {\bf 422} 681 \\ Sil A and Chatterjee S 1994 {\it Gen. Rel. Grav.}
 {\bf 26} 999
\bibitem{kk} Kaluza T 1921 {\it Sitz Preuss. Akad. Wiss.}
 {\bf D 33} 966 \\ Klein O 1926 {\it Z. Phys.} 895
\bibitem{js} Schwarz J J 1983 {\it Nucl. Phys.} {\bf B226} 269
\bibitem{he} Hawking S W and Ellis  G F R 1973 {\it The Large Scale Structure
of Space-time} (Cambridge University Press, Cambridge)
\bibitem{ft1} Tipler F J  1987 {\it Phys. Lett.} {\bf A 64} 8
\bibitem{ft} Tipler F J,  Clarke C J S  and Ellis G F R 1980
{\it General Relativity and Gravitation}, ed by A Held
(Plenum, New York)
\bibitem{ck} Clarke C J S and  Kr\'{o}lak A 1986 {\it J. Geom. Phys.} {\bf 2}
127
\end{thebibliography}
\end{document}